\documentclass[useAMS]{mn2e}
\usepackage{times}
\title{Thermal physics, cloud geometry, and the stellar IMF}

\author[Richard B. Larson]
       {Richard B. Larson\thanks{E-mail: larson@astro.yale.edu} \\
  Yale Astronomy Department, Box 208101, New Haven, CT 06520-8101, USA}

\date{Submitted 2004 December 15; Accepted 2005 January 31}

\pagerange{\pageref{firstpage}--\pageref{lastpage}}
\pubyear{2005}
\begin{document}
\maketitle
\label{firstpage}

\begin{abstract}

   The thermal properties of star-forming clouds have an important influence on how they fragment into stars, and it is suggested in this paper that the low-mass stellar IMF, which appears to be almost universal, is determined largely by the thermal physics of these clouds.  In particular, it is suggested that the characteristic stellar mass, a little below one solar mass, is determined by the transition from an initial cooling phase of collapse to a later phase of slowly rising temperature that occurs when the gas becomes thermally coupled to the dust.  Numerical simulations support the hypothesis that the Jeans mass at this transition point plays an important role in determining the peak mass of the IMF.  A filamentary geometry may also play a key role in the fragmentation process because the isothermal case is a critical one for the collapse of a cylinder: the collapse and fragmentation of a cylinder can continue freely as long as the temperature continues to decrease, but not if it begins to increase.  The limited available results on the dependence of the thermal properties of clouds on metallicity do not suggest a strong dependence of the IMF on metallicity, but the far-infrared background radiation in starburst regions and in the early universe may significantly shift the peak mass to higher masses in these situations.

\end{abstract}


\section{INTRODUCTION}  

   The problem of understanding the distribution of masses with which stars are formed, or the stellar initial mass function (IMF), remains unsolved.  However, observations have made much progress in recent years in clarifying the main features of the IMF that need to be understood.  Much evidence now supports a general form of the IMF that is similar to the original Salpeter (1955) power law at masses above one solar mass but flattens below one solar mass and declines below 0.1 solar masses when expressed in terms of the number of stars per unit logarithmic mass interval.  The IMF in logarithmic units is thus a peaked function, peaking at a few tenths of a solar mass (Miller \& Scalo 1979; Scalo 1986, 1998; Kroupa 2001, 2002; Chabrier 2003).  This means that nature makes stars with a preferred mass of this order.  The amount of mass that goes into stars in each logarithmic mass interval is also a peaked function that peaks at about 0.5 solar masses, according to the approximation suggested by Kroupa (2002).  Thus, in terms of where most of the mass goes, there is a characteristic stellar mass of the order of half of a solar mass.  This is perhaps the most fundamental fact about star formation that needs to be explained by any theory: some feature of the physics of star formation must yield a characteristic stellar mass that is a little less than one solar mass.

   A further remarkable fact is that the IMF shows a considerable degree of universality; similar and often indistinguishable results have been found in many different star-forming environments in our Galaxy and others, and no clear dependence has been found on any plausibly relevant astrophysical parameter such as metallicity.  Thus, not only must some feature of the physics of star formation lead to a characteristic mass scale, it must operate in a relatively universal way that depends only weakly on the environment and most astrophysical parameters.  This fact poses a further challenge to any theoretical understanding of star formation.

   One possibility for explaining a universal mass scale might be in terms of some universality in the properties of star-forming clouds or the initial conditions for star formation.  For example, if prestellar cloud cores are created by turbulent compression (Larson 1981; Mac Low \& Klessen 2004), a characteristic turbulent pressure might translate into a characteristic mass scale for star formation; the inferred turbulent pressures in nearby star-forming clouds are of the right magnitude for such an explanation to seem feasible (Larson 1996, 2003).  Alternatively, a characteristic magnetic field strength might also imply a mass scale for star formation, with magnetic pressure taking the place of turbulent pressure; our limited knowledge of magnetic field strengths in molecular cloud cores seems consistent with this hypothesis as well (Shu, Li \& Allen 2004).  However, star-forming environments vary so widely in their properties, and turbulence and magnetic fields are in any case such variable and poorly defined phenomena, that explanations based on turbulence or magnetic fields do not clearly offer a compelling explanation for a universal mass scale.

   Another possibility is that a characteristic stellar mass scale might arise from some universality in the basic physics of collapsing clouds, or in the physics of forming stars, that yields a mass scale depending only on fundamental physics.  For example, it has been suggested that protostellar accretion might be terminated at a characteristic stellar mass by feedback effects such as outflows that depend only on internal stellar processes such as deuterium burning (Shu, Adams \& Lizano 1987; Adams \& Fatuzzo 1996; Meyer et al.\ 2000).  While outflows probably do play some role in limiting protostellar accretion, theories of the IMF based on this hypothesis tend to contain many uncertain parameters, and this makes the importance of such effects very uncertain.

   In this paper, we address another aspect of the physics of star formation that has so far received relatively little attention in this context, namely the detailed thermal physics of collapsing and fragmenting clouds.  High-resolution mapping of star-forming clouds at millimeter wavelengths has suggested that low-mass stars form from, and acquire their masses from, observed dense prestellar cloud cores that have a mass spectrum resembling the stellar IMF (Andr\'e, Ward-Thompson \& Barsony 2000; Motte \& Andr\'e 2001).  If low-mass stars indeed derive their masses from those of the observed prestellar cores, the processes of cloud fragmentation must play an important role in determining the IMF.  The scale of cloud fragmentation is expected to depend on the Jeans mass, which in turn depends strongly on the temperature, and recent numerical simulations have shown that the amount of fragmentation that occurs is very sensitive to the exact temperature-density relation in collapsing clouds (Li, Klessen \& Mac Low 2003).  We discuss further in Section~2 the likely importance of fragmentation scales such as the Jeans length and mass, and in Section~3 we discuss the key role of the temperature-density relation in determining the form of the IMF and the characteristic stellar mass.  The thermal physics of star-forming clouds and the predicted temperature-density relation are reviewed in Section~4 on the basis of the existing literature, and the role of the geometry of collapsing clouds is emphasized in Section~5.  Section~6 discusses the dependence of the results on parameters such as metallicity and ambient far-infrared radiation in starburst regions and the early universe, with the implication that a top-heavy IMF may be plausible in the latter situations.  A summary of the main conclusions is presented in Section~7.

\section{THE SCALE OF CLOUD FRAGMENTATION}  

   Much evidence now indicates that low-mass stars form in small dense prestellar `cores' in star-forming molecular clouds; these cloud cores have approximately stellar masses, and they are often closely associated with newly formed stars (Myers 1985, 1999; Benson \& Myers 1989; Lada, Strom \& Myers 1993; Evans 1999; Williams, Blitz \& McKee 2000; Ward-Thompson 2002).  This close association suggests that these cores collapse directly to form stars, perhaps typically in binary or small multiple systems.  Several detailed mapping studies have found that the prestellar cores have a mass spectrum that resembles the stellar IMF, and it has been suggested on this basis that low-mass stars acquire their masses directly from those of the observed cores (Motte, Andr\'e \& Neri 1998; Testi \& Sargent 1998; Luhman \& Rieke 1999; Motte et al.\ 2001; Motte \& Andr\'e 2001).  Other similar studies have yielded core masses that are systematically somewhat larger (Johnstone et al.\ 2000, 2001), but the resulting mass spectrum again has a shape similar to the stellar IMF, and it is still consistent with the direct collapse of these cores into stars with a somewhat lower efficiency.  All studies agree that the masses of the cores are similar in order of magnitude to the masses of low-mass stars.

   If low-mass stars form by the direct collapse of prestellar cloud cores and acquire masses similar to those of the cores, this suggests that the characteristic stellar mass is determined largely by the processes of cloud fragmentation.  Understanding the origin of stellar masses then requires understanding how the material in star-forming clouds becomes divided up into individual star-forming units.  The clearest evidence that the characteristic stellar mass depends mainly on the scale of cloud fragmentation may be provided by the fact that most stars form in small clusters containing a few hundred stars where the efficiency of star formation is moderately high, of the order of 25 to 30 percent (Lada \& Lada 2003).  This means that the average stellar mass, which is just the total mass of the stars in each region divided by the number of stars present, is determined to within a factor of 3 or 4 just by the number of stars formed in each region.  The problem of understanding the characteristic stellar mass then becomes basically a problem of understanding the number of prestellar cores that form in each region.

   What controls the scale of cloud fragmentation, or the number of objects formed?  Classically, this has been thought to be the balance between thermal pressure and gravity, and a minimum length scale $\lambda_{\rm J}$ for growing density fluctuations in a medium with a finite pressure was first derived by Jeans (1902, 1929).  A corresponding minimum mass or Jeans mass $M_{\rm J}$ can be derived from the Jeans length by assuming a geometry for the growing density fluctuations (Spitzer 1978; Larson 1985, 2003).  Here we adopt for convenience the Jeans mass in the usual form given by Spitzer (1978), which is just the mass in a cube whose side is the Jeans length.  Although the analysis of Jeans was inconsistent in that it neglected the overall collapse of the background medium, rigorous stability analyses have been made for equilibrium configurations such as sheets, filaments, and spheres, and have yielded dimensionally equivalent results (Larson 1985, 2003).  For example, if spherical symmetry is assumed and it is supposed that collapse begins with a marginally stable isothermal sphere or `Bonnor-Ebert sphere', then the mass of such a sphere is dimensionally the same as the Jeans mass, but it is numerically smaller by a factor of 4.7 because a Bonnor-Ebert sphere contains only matter whose density is above the background density, while a region one Jeans length across also contains matter of lower density that may or may not collapse with the denser gas (Larson 2003).  In numerical simulations of cloud collapse and fragmentation, the number of fragments that form is generally comparable to the number of Jeans masses present initially (Larson 1978; Monaghan \& Lattanzio 1991; Klessen, Burkert \& Bate 1998; Klessen 2001; Bate, Bonnell \& Bromm 2003); in fact, Bate \& Bonnell (2005) find that the median fragment mass scales closely with the initial Jeans mass.  These results suggest that the Jeans mass may be of quite general importance in determining the scale of cloud fragmentation, at least whenever thermal pressure is the most important force resisting gravity.

   Observationally, it has been argued that the small but apparently real differences in the mass spectra of the young stars in different nearby regions of star formation can be understood if the peak mass of the IMF depends on the Jeans mass in the associated clouds.  In particular, denser clouds in which the Jeans mass is smaller appear to produce stars with a mass spectrum that peaks at a lower mass and contains more brown dwarfs, qualitatively as expected if the Jeans mass plays an important role in cloud fragmentation (Brice\~no et al.\ 2002; Luhman et al.\ 2003; Luhman 2004).

   The relevance of the Jeans mass has, however, been much debated because effects such as turbulence and magnetic fields are expected to be much more important than thermal pressure in resisting gravity during the early stages of cloud evolution (Mac Low \& Klessen 2004).  The angular momentum of star-forming cores has also been thought to present a serious obstacle to star formation (Larson 2002).  However, turbulence is predicted to decay rapidly as a cloud evolves, even if it consists predominantly of MHD waves (Mac Low \& Klessen 2004).  Dissipation of turbulence may even be required to allow the gas to condense into bound prestellar cores (Nakano 1998; Myers \& Lazarian 1998); the turbulence in prestellar cores is in fact observed to be subsonic and therefore less important than thermal pressure in supporting them against gravity (Myers 1983; Goodman et al.\ 1998).  The structures of many cores appear to be adequately approximated as thermally supported Bonnor-Ebert spheres (Williams et al.\ 2000; Ward-Thompson 2002.)  Rotation is also too small an effect to influence significantly the early stages of core collapse (Goodman et al.\ 1993), and its main effect on the later stages is probably just the formation of binary or small multiple systems (Larson 2002).  Numerical simulations of turbulent collapsing clouds have shown that the number of bound clumps that form is not very sensitive to the way in which the turbulence is treated, or even to whether turbulence is initially present at all; the number of bound clumps formed is always comparable to the number of Jeans masses present initially, although fragment masses may be somewhat reduced by compression during the collapse (Klessen 2001; Bonnell \& Bate 2002; Bate et al.\ 2003.)

   The effect of a magnetic field on cloud fragmentation has been less well explored, but simulations suggest that the nature of the turbulence in star-forming clouds is not greatly altered by the presence of a magnetic field, since similar filamentary and clumpy structures are still found (Ostriker, Gammie \& Stone 1999; Mac Low \& Klessen 2004; Li et al.\ 2004).  A few limited simulations of fragmentation that have included both turbulence and magnetic fields suggest that the presence of a magnetic field may reduce the efficiency of star formation, but they do not show a clear effect of the magnetic field on the mass spectrum of the objects formed (Li et al.\ 2004; V\'azquez-Semadeni et al.\ 2005).  In any case, star-forming cloud cores are expected to lose much of their magnetic flux at a relatively early stage by ambipolar diffusion, a process that may be accelerated by turbulence (Heitsch et al.\ 2004).  Nakano (1998) has argued that the observed prestellar cores cannot be predominantly magnetically supported because they would then not show their observed large enhancements in column density.  Direct measurements of the magnetic field strengths in cloud cores are difficult and have mostly yielded only upper limits, but the available data are consistent with a level of magnetic support that may be comparable to but not greater than thermal pressure in importance (Crutcher 1999; Bourke et al.\ 2001).  Furthermore, models of magnetized cloud cores fit the observations best if they have magnetic fields that are comparable to but not dominant over thermal pressure in importance (Ciolek \& Basu 2000, 2001).  If the magnetic field is not strong enough to completely suppress gravitational instability, the minimum scale for fragmentation is not very different from the Jeans scale (Larson 1985), and the collapse of a magnetized cloud core proceeds in a way that is qualitatively similar to the non-magnetic case but is somewhat slowed by the magnetic field (Basu 1997, 1998; Larson 2003).

   In summary, both simulations and observations suggest that nature eventually disposes of most of the initial turbulence and magnetic fields in star-forming clouds to form prestellar cores in which thermal pressure provides the dominant support against gravity.  Turbulence eventually becomes unimportant because it is dissipative and is predicted to decay rapidly even if it is dominated by MHD waves.  Magnetic fields can be dissipated by ambipolar diffusion and field reconnection, and magnetic forces are in any case anisotropic and cannot prevent matter from accumulating along the field lines.  By contrast, thermal pressure cannot be dissipated or reduced below a minimum set by the lowest attainable temperature; moreover, as an isotropic force, it cannot be overcome by any anisotropic motions.  Thermal pressure is thus a final irreducible barrier to star formation that remains even after turbulence and magnetic fields have been dissipated, and that cannot be overcome except by the accumulation of at least a Jeans mass of material.  The Jeans scale must therefore play a key role in at least the final stages of the star formation process, regardless of what happens during the earlier evolution of star-forming clouds.

\section{IMPORTANCE OF THE THERMAL PHYSICS}  

   Most numerical simulations of cloud collapse and fragmentation have assumed that the gas remains isothermal during the early stages of collapse; this approximation has been based on early studies of the thermal behavior of star-forming clouds which had found that the temperature does not change much over the relevant range in density (McNally 1964; Hayashi \& Nakano 1965; Hayashi 1966; Hattori, Nakano \& Hayashi 1969; Larson 1973).  As was noted above, simulations of isothermal collapse and fragmentation have shown that the number of bound fragments formed is generally similar to the number of Jeans masses present initially (e.g., Larson 1978; Monaghan \& Lattanzio 1991; Klessen et al.\ 1998; Klessen 2001; Bate et al.\ 2003; Bate \& Bonnell 2005).  The Jeans mass depends strongly on the temperature, varying either as $T^{3/2}\rho^{-1/2}$ or as $T^2P^{-1/2}$ according to whether the density or the pressure is specified, and therefore the thermal properties of collapsing clouds must play an important role in determining the scale of fragmentation.  The usual isothermal approximation is actually only a fairly crude one and may not be adequate, since the above studies found that the temperature in collapsing clouds can vary by as much as a factor of 2 or more above and below the usually assumed constant value of 10~K.  If the temperature varies significantly during the collapse, this might have an important effect on the amount of fragmentation that occurs; for example, Monaghan \& Lattanzio (1991), Turner et al.\ (1995), and Whitworth et al.\ (1995) found that fragmentation is greatly enhanced if significant cooling occurs during the collapse.

   Larson (1985) suggested that the detailed thermal behavior of a fragmenting cloud might play a key role in determining the extent to which fragmentation can continue.  As was shown in that paper, the temperature is predicted to decrease significantly with increasing density at low densities, causing the Jeans mass to decrease strongly as the density rises; this suggests that fragmentation might be strongly favored at these low densities.  At high densities, however, the temperature is predicted to increase slowly with increasing density, causing the Jeans mass to decrease much more slowly, and this suggests that continuing fragmentation might be much less likely at these higher densities.  The density at which the temperature reaches its minimum value and begins to increase may therefore be the highest density at which fragmentation is probable, and it may imply a preferred mass scale for fragmentation.  Larson (1985, 1986) suggested that this preferred mass scale might lead to a peak in the IMF at about this mass, and noted that the Jeans mass at the temperature minimum is predicted to be about 0.3 solar masses, similar to the mass at which the observed IMF peaks.  The temperature minimum occurs when the gas becomes thermally coupled to the dust, and the preferred mass scale may therefore depend on the conditions under which the gas and dust become thermally coupled.  The possibility that the onset of thermal coupling between gas and dust might result in a preferred scale for fragmentation was previously suggested by Whitworth, Boffin \& Francis (1998).

   The work of Li et al.\ (2003) shows that the amount of fragmentation that occurs is indeed very sensitive to the assumed temperature-density relation.  These authors simulated the collapse and fragmentation of turbulent clouds with simple polytropic equations of state of the form $P \propto \rho^\gamma$, corresponding to a temperature-density relation of the form $T \propto \rho^{(\gamma - 1)}$.  A value of $\gamma < 1$ thus corresponds to cooling during the collapse, while $\gamma > 1$ corresponds to heating.  Li et al.\ (2003) found that the number of fragments that forms depends strongly on $\gamma$, especially for values near unity; for example, a simulation with $\gamma = 0.7$ produced about 380 bound clumps, while an otherwise identical one with $\gamma = 1.1$ produced only about 18 bound clumps, a factor of 20 fewer.  These values of $\gamma$ are relevant to star formation because they are approximately the values that are expected to apply at low and high densities in star-forming clouds: at low densities $\gamma \simeq 0.73$ (Larson 1985), while at high densities $\gamma \simeq 1.07$ (Section~4.)  It might then be expected that much more fragmentation would occur at the low densities where $\gamma \simeq 0.73$ than at the high densities where $\gamma \simeq 1.07$, and this should have important consequences for the stellar IMF, favoring masses similar to the Jeans mass at the transition point where $\gamma$ increases from $\gamma < 1$ to $\gamma > 1$.  The characteristic stellar mass might then be determined largely by the detailed thermal physics of star-forming clouds. 

   An example of a case where the thermal physics has a clear impact on the scale of fragmentation is provided by studies of the formation of the first stars in the universe.  In the absence of heavy elements, the thermal physics of the first star-forming clouds is relatively simple and is controlled by trace amounts of molecular hydrogen.  As reviewed by Bromm \& Larson (2004), several groups have made detailed calculations of the collapse of the first star-forming clouds and have arrived at a consistent picture of their thermal behavior (Abel, Bryan \& Norman 2002; Bromm, Coppi \& Larson 2002; Nakamura \& Umemura 2002).  After an initial phase of rapid collapse and heating, cooling by line emission from H$_2$ molecules reduces the temperature to a minimum value of around 200~K set by the level spacing of the H$_2$ molecules.  When the density rises above about $10^4$\,cm$^{-3}$, cooling becomes less efficient because the upper levels of the molecules become thermalized; the temperature then begins to rise slowly with increasing density, causing the contraction of the star-forming clumps to decelerate.  The Jeans mass at this point is of the order of $10^3$\,M$_\odot$, and the numerical simulations all yield clump masses of the order of several hundred to a thousand solar masses.  A parameter study by Bromm et al.\ (2002) showed that this mass scale is relatively robust and depends mainly on the physics of the H$_2$ molecule, i.e.\ on fundamental atomic and molecular physics, with only weak dependences on other parameters such as the cosmological initial conditions.  Although the final evolution of these first star-forming clumps has not yet been determined, it seems likely that the result will be the formation of a very massive star whose mass may be 100~M$_\odot$ or more (Bromm \& Larson 2004).

   In present-day clouds, a qualitatively similar situation is expected to occur because at low densities, efficient atomic cooling at a rate per unit mass that increases with density causes the temperature to decrease with increasing density, while at high densities the cooling rate per unit mass no longer increases with density and the temperature begins to rise slowly (see Section~4.)  Again, it might be expected that the Jeans mass at the temperature minimum would be a preferred scale of fragmentation.  If the predicted variation of temperature with density in collapsing clouds is indeed important in determining the stellar IMF, it is then necessary to understand as well as possible the detailed thermal physics of star-forming clouds; this subject will be discussed further in the following section where the relevant literature will be reviewed.

   Another mass scale in star formation that depends only on fundamental physics is the `opacity-limit' mass, a lower limit on the mass scale for fragmentation that is determined by the onset at high density of a high opacity to the thermal emission from dust (Low \& Lynden-Bell 1976).  Low \& Lynden-Bell estimated that the opacity-limit mass is about 0.007 solar masses, almost two orders of magnitude smaller than the characteristic stellar mass mentioned above in Section~1.  Thus, the opacity limit is evidently not what determines the characteristic stellar mass or accounts for the peak in the IMF, but it may nevertheless set a lower limit on stellar masses, or more precisely on the masses of objects that form like stars by independent fragmentation, as distinct from objects that form like planets in protostellar disks (although this distinction may not be entirely sharp.)  Thus there may be two fundamental mass scales in star formation, a characteristic mass and a minimum mass, both of which are determined by the thermal physics.  The characteristic mass may reflect the Jeans mass at the point where the cooling rate per unit mass stops increasing and the temperature reaches a minimum, while the minimum mass reflects the Jeans mass at the point where cooling is completely suppressed by opacity and the collapse becomes adiabatic.

\section{THERMAL PROPERTIES OF STAR-FORMING CLOUDS}  

   Observations and theory both indicate that there are two basic thermal regimes in star-forming clouds (Larson 1973, 1985): (1) At low densities, i.e.\ below about $10^{-19}$\,g\,cm$^{-3}$, the temperature decreases strongly with increasing density, controlled by a balance between radiative heating and cooling processes; heating is due mainly to photoelectric emission from dust grains, possibly with a contribution from cosmic rays, while cooling is due to collisionally excited line emission from common atoms and ions such as C$^+$ and O.  Because the associated cooling rate per unit mass increases with density, the predicted temperature decreases with increasing density, eventually falling below 10~K at densities above $10^{-19}$\,g\,cm$^{-3}$.  (2) At higher densities, cloud cores become optically thick to the heating radiation and atomic line emission that are important at low densities, and these processes become unimportant.  Cooling by molecules may play some role at intermediate densities, but this role is limited because most of the molecules soon freeze out onto the dust grains (Goldsmith 2001).  At densities above about $10^{-18}$\,g\,cm$^{-3}$ the gas eventually becomes thermally coupled to the dust, which subsequently controls the temperature by its far-infrared thermal emission (see also Whitworth et al.\ 1998).  At these high densities, heating is expected to be due mainly to compression at nearly the free-fall rate; since the associated heating rate increases with increasing density, the temperature begins to increase slowly with density.  Figure 2 of Larson (1985) summarizes the observational and theoretical results then available, and it shows that in the low-density regime, theory and observations are in good agreement and indicate a similar trend of decreasing temperature with increasing density.  At higher densities, the gas temperature becomes more difficult to measure because most of the molecules used for this purpose freeze out onto the grains, and it becomes necessary to rely mainly on theory for temperature estimates.  The theoretical prediction shown in Figure 2 of Larson (1985), taken from Larson (1973), shows the temperature reaching a minimum value of 5~K at a density of $2 \times 10^{-18}$\,g\,cm$^{-3}$, after which it begins to rise slowly with increasing density.

   Koyama \& Inutsuka (2000) have studied again the thermal properties of star-forming clouds in the low-density regime, assuming that the heating is due primarily to the photoelectric effect rather than to cosmic rays, as had been assumed in most previous work.  Their predicted temperature-density relation is similar to those shown by Larson (1985) for densities up to about $2 \times 10^{-20}$\,g\,cm$^{-3}$, while yielding somewhat higher temperatures at higher densities.  At these higher densities, predictions become more uncertain for several reasons: optical depth and radiative transfer effects become important, and cooling by molecules and small dust grains may also become important.  Observations show that low-mass prestellar cores with such densities generally have temperatures around 10~K, with some variation between different regions of star formation (Myers 1985, 1999; Benson \& Myers 1989; Evans 1999).  As was reviewed by Evans (1999), the typical temperature of a dense low-mass core with a density of $10^{-19}$\,g\,cm$^{-3}$ is about 8.5~K, and this is consistent with a smooth extension of the temperature-density relation seen at the lower densities, and with the continuing validity of the polytropic approximation with $\gamma \simeq 0.73$ suggested by Larson (1985) up to a density of at least $10^{-19}$\,g\,cm$^{-3}$.

   It is unclear observationally whether the temperature continues to decrease to even lower values at higher densities, as is predicted by most of the theoretical calculations.  Larson (1985) and Masunaga \& Inutsuka (2000) both predicted a minimum temperature of 5~K at a density of 2 to $3 \times 10^{-18}$\,g\,cm$^{-3}$, while cloud models that include radiative heating of the dust by ambient far-infrared radiation predict a somewhat higher minimum temperature of about 6 to 7~K at a similar density (Zucconi, Walmsley \& Galli 2001;  Evans et al.\ 2001; Galli, Walmsley \& Gon\c calves 2002).  Observations have generally indicated still higher temperatures of at least 8~K and typically about 10~K (Benson \& Myers 1989; Ward-Thompson, Andr\'e \& Kirk 2002), even in the densest cloud cores (Tafalla et al.\ 2004).  Thus the temperature may not in reality become quite as low as is predicted theoretically, although the possibility has not been excluded that the densest cloud cores contain very cold gas that has not yet been detected.

   At the highest densities, where it is necessary to rely entirely on theory for temperature estimates, a number of calculations have predicted very similar temperature-density relations (Larson 1973; Low \& Lynden-Bell 1976; Masunaga \& Inutsuka 2000; Omukai 2000).  In all cases it has been assumed that heating is caused by dynamical compression at nearly the free-fall rate, and that cooling is due to far-infrared thermal emission from the dust.  Because of the strong temperature dependence of the dust emission, which varies as the 6th or 7th power of the temperature, the resulting temperature depends only weakly on the density or on the grain properties.  For densities between $10^{-18}$\,g\,cm$^{-3}$ and $10^{-13}$\,g\,cm$^{-3}$, the first three of the above studies (Larson 1973; Low \& Lynden-Bell 1976; Masunaga \& Inutsuka 2000) agree to within 0.1~dex in predicting temperatures that increase slowly from about 5~K to 13~K over this density range, while the fourth (Omukai 2000) predicts temperatures that are systematically about 25 percent lower.  These calculations did not include radiative heating of the dust by ambient far-infrared radiation, which is likely to raise the minimum temperature to at least 6 to 7~K, as was noted above.  Radiative heating of the dust becomes even more important in regions of active star formation such as starburst regions, as will be discussed further in Section~6.2.

   Most of the above calculations did not incorporate a distribution of grain sizes, but assumed grains of a single size.  At high densities where the gas is thermally well coupled to the dust, only the total mass in dust is relevant for the thermal balance and the size distribution is unimportant, but at intermediate densities the size distribution can be important because it affects the collision rate between gas and dust and hence the density at which the gas and dust become thermally coupled.  Among the studies mentioned above, only Omukai (2000) incorporated a realistic distribution of grain sizes, and as was noted above he found lower temperatures, including an unrealistically low minimum temperature of about 2.6~K at a density of $2 \times 10^{-19}$\,g\,cm$^{-3}$.  As another illustration of the possible effect of smaller grain sizes, the temperatures predicted by Larson (1973) were recalculated assuming the same dust mass but grain sizes 10 times smaller; the resulting temperatures are unchanged at the lowest and highest densities, but reach a lower minimum of 4~K at a density of $2 \times 10^{-19}$\,g\,cm$^{-3}$.  Again, however, such low temperatures seem unlikely to be realistic because of radiative heating of the dust by ambient far-infrared radiation.  Since the uncertainties are particularly large at these intermediate densities, it may be better for approximate purposes to use interpolations based on the simple power-law fits that work well at the lower and higher densities.

   Within the uncertainties, the observational and theoretical results discussed above for the low- and high-density regimes can be represented by the following approximate temperature-density relation consisting of two power laws:
\[
   T=4.4\,\rho_{18}^{-0.27}\,{\rm K},~~\rho < 10^{-18} {\rm g\,cm}^{-3}
\]
\[
   T=4.4\,\rho_{18}^{+0.07}\,{\rm K},~~\rho > 10^{-18} {\rm g\,cm}^{-3}
\]
\noindent
where $\rho_{18}$ is the density in units of $10^{-18}$\,g\,cm$^{-3}$.  This approximation to the equation of state, in which $\gamma$ changes from 0.73 to 1.07 at a density of $10^{-18}$\,g\,cm$^{-3}$, is valid to within about $\pm 0.1$~dex for most densities between $10^{-22}$\,g\,cm$^{-3}$ and $10^{-13}$\,g\,cm$^{-3}$, except that at densities between $3 \times 10^{-20}$\,g\,cm$^{-3}$ and $3 \times 10^{-17}$\,g\,cm$^{-3}$ the uncertainty is larger, perhaps $\pm 0.2$~dex.  The actual temperature minimum will be somewhat smoothed out compared with this simple two-part approximation, especially if radiative heating of the grains is important and raises the minimum temperature significantly.  More detailed treatments of the thermal physics and better observational constraints will be needed to obtain more definitive results.

   The details may, in any case, not be as important as the basic fact that the temperature in star-forming clouds is expected to decrease strongly with increasing density in the low-density regime, i.e.\ at densities below about $10^{-19}$\,g\,cm$^{-3}$, while it is expected to vary much less in the high-density regime, at least until the opacity limit is reached at a density of $10^{-13}$\,g\,cm$^{-3}$.  According to Li et al.\ (2003), the largest drop in the efficiency of fragmentation occurs when $\gamma$ is increased from 0.7 to~1.0; the number of fragments decreases by a factor of 7 over this range of $\gamma$, but only by another factor of 3 when $\gamma$ is increased further to~1.1.  Thus, the most important feature of the thermal behavior of star-forming clouds may just be the change from an initial cooling phase of collapse to a later nearly isothermal phase.  Even if the temperature never drops any lower than, say, 8~K, which is approximately the lowest temperature directly inferred from observations, and if it stays approximately constant at higher densities, the Jeans mass at the transition point between the cooling phase and the nearly isothermal phase should still be an important mass scale for fragmentation.  According to the approximation given above, the temperature in the low-density regime falls to 8~K at a density of about $10^{-19}$\,g\,cm$^{-3}$, at which point the Jeans mass is about 2~M$_\odot$ while the Bonnor-Ebert mass, another commonly used mass scale, is about 0.5~M$_\odot$.  These numbers are still of the right magnitude to be relevant to the characteristic stellar mass.

   To explore numerically the effect of an equation of state in which $\gamma$ changes from 0.7 at low densities to 1.1 at high densities, Jappsen et al.\ (2005; see also Klessen et al.\ 2005) have made detailed simulations of cloud collapse and fragmentation that are similar to those of Li et al.\ (2003) except that $\gamma$ is assumed to change from 0.7 to 1.1 at some critical density $\rho_{\rm crit}$.  The value of $\rho_{\rm crit}$ is then varied to test the effect of this parameter on the mass spectrum of the resulting bound clumps.  The results of Li et al.\ (2003) and the discussion in Section~3 suggest that the Jeans mass at the point of minimum temperature should be a  preferred scale for fragmentation, and that it may determine the peak mass of the IMF; if this is correct, the typical mass of the bound clumps that form should scale approximately with the Jeans mass $M_{\rm J,crit}$ at the density $\rho_{\rm crit}$ where the temperature reaches its minimum value and the gas becomes thermally coupled to the dust (see also Whitworth et al.\ 1998).  The results of Jappsen et al.\ (2005) show a clear dependence of the clump mass spectrum on $\rho_{\rm crit}$ in the expected sense that as $\rho_{\rm crit}$ increases and $M_{\rm J,crit}$ decreases, the number of bound clumps increases and their median mass decreases.  The median mass of the clumps is also typically similar to the Jeans mass $M_{\rm J,crit}$ at the critical density.  These results confirm that the thermal properties of star-forming clouds play an important role in determining the stellar IMF, and they add support to the hypothesis that the Jeans mass at the point of minimum temperature is important in determining the characteristic stellar mass.

\section{ROLE OF THE GEOMETRY OF FRAGMENTING CLOUDS}  

   A striking feature of the simulations of Jappsen et al.\ (2005) is the prominence of filaments and the fact that nearly all of the bound objects form in the filaments.  Filamentary structure is a very common feature of simulations of collapse and fragmentation, appearing in many simulations regardless of whether turbulence or magnetic fields are included (e.g., Monaghan \& Lattanzio 1991; Klessen 2001; Klessen \& Burkert 2001; Bonnell \& Bate 2002; Bate et al.\ 2003; Klessen et al.\ 2004; Li et al.\ 2003, 2004; Padoan et al.\ 2005).  Filamentary structure is also seen in many observed star-forming clouds, and much of the star formation in these clouds appears to occur in filaments (Schneider \& Elmegreen 1979; Larson 1985; Curry 2002; Hartmann 2002.)  Many star-forming cloud cores are elongated and probably prolate, sometimes clearly forming parts of larger elongated or filamentary structures (Myers et al.\ 1991).  Thus, the formation and fragmentation of filaments may be an important aspect of star formation quite generally (Larson 1992).  If most of the fragmentation that forms stars occurs in filaments, this may help to explain the result of Li et al.\ (2003) that the amount of fragmentation depends strongly on the value of $\gamma$, especially for values of $\gamma$ near unity.  This is because the isothermal case $\gamma = 1$ is a critical one for the collapse of a cylinder: if $\gamma < 1$, a cylinder can collapse indefinitely toward its axis and fragment indefinitely into very small objects, while if $\gamma > 1$, this is not possible because pressure then increases faster than gravity and the collapse is halted at a finite maximum density where the Jeans mass has a finite minimum value (Mestel 1965).  Fragmentation should therefore be favored when $\gamma < 1$ and suppressed when $\gamma > 1$, as was indeed found by Li et al.\ (2003).

   Insight into the dynamics of collapsing configurations can be obtained from similarity solutions, which can often approximate the behavior of systems with simple polytropic equations of state (Larson 2003).  Similarity solutions for collapsing cylinders with polytropic equations of state have been derived by Kawachi \& Hanawa (1998), and these authors found that the existence of such solutions depends on the value of $\gamma$: similarity solutions exist for $\gamma < 1$, but not for $\gamma > 1$.  For the solutions with $\gamma < 1$, the collapse becomes slower and slower as $\gamma$ approaches unity from below, asymptotically coming to a halt when $\gamma = 1$.  This result shows in a particularly clear way that $\gamma = 1$ is a critical case for the collapse of filaments.  Kawachi \& Hanawa (1998) suggested that the deceleration of the collapse that is predicted to occur as $\gamma$ approaches unity will cause a filament to fragment into clumps because the timescale for collapse toward the axis then becomes longer than the timescale for fragmentation.  If $\gamma$ increases with increasing density, as is expected from the discussion in Sections 3 and~4, the collapse of a filament toward its axis would be expected to decelerate as $\gamma$ approaches unity, favoring the breakup of the filament into clumps at this point.  The sensitivity of cylinder fragmentation to the equation of state, and the termination of fragmentation when $\gamma$ rises above unity, were also noted by Bate \& Burkert (1997) on the basis of numerical simulations.  The Jeans mass at the point where $\gamma$ rises above unity and the temperature reaches a minimum is predicted to be about 0.3 solar masses (Section~3), similar to the mass at which the IMF peaks.  This similarity suggests that the collapse and fragmentation of filaments with an increasing value of $\gamma$ may play a central role in the origin of the stellar IMF.

   The isothermal equation of state is a special case also for another well-studied collapse problem, the collapse of a uniformly rotating cloud to a centrifugally supported disk.  If a uniform and uniformly rotating cloud collapses isothermally with conservation of angular momentum to form a centrifugally supported disk, the result is the singular isothermal disk studied by Toomre (1982) and Hayashi, Narita \& Miyama (1982), which can be regarded as a generalization to finite temperature and thickness of the infinitely thin disk first studied by Mestel (1963).  The singular isothermal disk, which has been generalized further to the magnetic case by Li \& Shu (1996), is a self-similar structure in which the rotation speed is constant everywhere and the Toomre stability parameter $Q$ is also constant everywhere.  For a realistic initial rotation rate the disk has $Q < 1$ when fully formed, and it should therefore be unstable to fragmentation, or at best only marginally stable (Larson 1984).  Detailed calculations of isothermal collapse with rotation have shown that a singular disk is indeed formed, and that it is built up from the inside out (Matsumoto, Hanawa \& Nakamura 1997; Saigo \& Hanawa 1998).  Instability should therefore occur first in the inner part of the disk, where it may result in the formation of a binary or multiple system.

   A self-similar disk is possible only for $\gamma = 1$; if $\gamma \ne 1$ the self-similarity is lost and the value of $Q$ becomes a function of radius.  If $\gamma < 1$, the innermost part of the disk becomes very thin and unstable to fragmentation ($Q \ll 1$), whereas if $\gamma > 1$ the inner disk is strongly stabilized ($Q > 1$).  Fragmentation of the inner disk is therefore unavoidable if $\gamma < 1$, but is prevented if $\gamma > 1$.  Calculations of the collapse of rotating clouds with various values of $\gamma$ confirm that when $\gamma < 1$ the inner part of the disk becomes very thin and unstable, typically having $Q \sim 0.35$; in fact a torus then always forms near the center, and such a configuration is highly unstable to fragmentation (Saigo, Matsumoto \& Hanawa 2000).  When $\gamma > 1$, these calculations show that the inner part of the disk is thick and stable against fragmentation.  The isothermal case $\gamma = 1$ is thus a special marginal case between stability and instability, and this may explain why the isothermal collapse of a rotating cloud has been such a challenging problem numerically, with some calculations yielding rings and others not, depending on the numerical resolution used (Larson 2003).  Calculations of rotating collapse and fragmentation in which $\gamma$ is assumed to increase from 1 to 7/5 when the opacity limit is reached show that central collapse is halted at this point, and fragmentation into a binary or small multiple system is then likely to occur (Bonnell 1994; Bonnell \& Bate 1994a,b; Whitworth et al.\ 1995; Matsumoto \& Hanawa 2003).

   The isothermal case therefore plays a special role in rotating clouds as well as in filamentary ones.  In both cases, axial collapse is halted or slowed when $\gamma$ rises above unity, and the scale of fragmentation is likely to be largely defined at this point.  Rotation, however, is generally observed to be a small effect in star-forming clouds, and it becomes important only at high densities where it is likely to result only in the formation of binary or multiple systems (Larson 2002, 2003).  This means that the fragmentation of rotating configurations is probably only of secondary importance for the IMF, influencing stellar masses by factors of 2 or 3 but not in order of magnitude.  Filamentary structure may therefore still play the most important role in the fragmentation of star-forming clouds and the origin of stellar masses.

   What might account for the prevalence of filamentary structure in star-forming clouds?  One way to produce filaments is by purely dynamical phenomena such as shear and vorticity, and such effects might account for some of the wispy structure seen in simulations of turbulent star-forming clouds (Mac Low \& Klessen 2004) and perhaps for much of the filamentary structure seen in turbulent flows generally.  Purely gravitational effects can also create filamentary structure in self-gravitating clouds by amplifying departures from spherical symmetry during collapse; for example, a pressure-free oblate spheroid collapses to a disk and a prolate spheroid to a line (Lin, Mestel \& Shu 1965).  Even with a finite pressure, a prolate configuration can collapse isothermally to a thin filament if its line density exceeds a critical value (McCrea 1957; Mestel 1965; Larson 1972).  A related possibility is that a flattened configuration with an elongated or irregular boundary can collapse to form an irregular filamentary shape (Burkert \& Hartmann 2004).  Thus gravity alone can create filamentary structures.  This may be especially the case in cosmology, where simulations of the dynamics of the cold dark matter have shown the ubiquitous appearance of filamentary networks.

   Another possibility is that the equation of state may itself play a role in determining the geometry of collapse and fragmentation.  Spherical collapse can occur for any value of $\gamma$ less than 4/3, but spherical collapse is not likely to result in much fragmentation because it quickly becomes strongly centrally concentrated and tends to produce a single central density peak.  Even with the addition of a realistic amount of rotation, fragmentation probably still doesn't proceed beyond the formation of a binary or multiple system.  Extensive fragmentation can only occur if many centers of collapse develop, and this requires large departures from spherical geometry.  The growth of departures from spherical symmetry is however inhibited by the increase of pressure forces, by an amount that depends on the equation of state and the geometry.  For example, in the isothermal case the ratio of pressure to gravity decreases during spherical collapse, remains constant during cylindrical collapse, and increases during planar collapse.  Indefinite collapse to a infinitely thin sheet is therefore not possible in the isothermal case.  In fact, collapse to a thin sheet is not possible for any value of $\gamma > 0$.  Collapse to a thin filament is also prevented for $\gamma > 1$.  In the low-density regime where $\gamma \sim 0.7$, collapse to a thin sheet is therefore not possible but collapse to a thin filament remains possible, leaving the formation of thin filaments as the most promising allowed path to fragmentation.  In the high-density regime where $\gamma > 1$, even collapse to a thin filament is not possible, and the only remaining allowed mode of collapse is the nearly spherical collapse of clumps formed by the fragmentation of filaments at lower densities.  These considerations may help to explain the apparently ubiquitous role of filaments in the fragmentation of star-forming clouds.  In this view, fragmentation may be seen as involving an interplay between the thermal physics and the geometry of collapsing clouds.

   Although indefinite collapse to a thin sheet cannot occur if $\gamma > 0$, the gas in a cloud can still be compressed one-dimensionally to a thin sheet by external pressures produced by events such as a cloud collision or supernova shock.  Shock compression has often been postulated as a mechanism to trigger star formation.  However, as long as $\gamma > 0$, a dense layer created by external compression will not be self-gravitating and will not fragment gravitationally unless it is maintained in its compressed state for a sufficiently long time for gravitational instability to develop; this time can be much longer than the sound crossing time or the free-expansion time.  This implies, for example, that a cloud collision can trigger star formation only if it is relatively slow and has a sufficiently long timescale (Smith 1980).  But for a slow collision the compression factor is not very large, and this limits the effectiveness of collisions in promoting fragmentation.  Thus, even with the help of external compression, sheet formation does not appear a very promising path toward fragmentation and star formation.

   A special case of sheet fragmentation that might nevertheless be relevant for the formation of some low-mass objects is the fragmentation of circumstellar disks around forming stars.  Bate, Bonnell \& Bromm (2002) found that most of the `proto-brown-dwarfs' formed in their collapse simulations were created in circumstellar disks by the fragmentation of thin spiral filaments that formed in these disks.  A circumstellar disk can avoid the time-scale limitation mentioned above because it does not have to be self-gravitating to survive, being confined by the gravity of the central star.  This allows the disk to grow stably in mass by accreting material until it becomes unstable to fragmentation, at which point it may form one or more small companion objects that could be either small stars or large planets.  A companion object formed in this way might sometimes grow into a binary stellar companion of similar mass, helping to solve the angular momentum problem (Larson 2002).  Such mechanisms may help to populate the low-mass end of the stellar IMF, but they do not seem likely to change much the overall form or mass scale of the IMF.

   A generalization of the simple cylindrical geometry discussed above would be a more complex fractal-like filamentary network with a higher dimension; many numerical simulations have indeed shown the formation of filamentary networks (e.g., Monaghan \& Lattanzio 1991; Klessen et al.\ 1998; Klessen \& Burkert 2001; Bonnell \& Bate 2002; Jappsen et al.\ 2005).  A cloud model based on a fractal filamentary network was suggested by Larson (1992).  If a fractal filamentary network forms in a collapsing cloud, its dimension may depend on the value of $\gamma$: smaller values of $\gamma$ may lead to more complex space-filling networks of higher dimension because the stronger associated cooling may allow more small-scale structure and more branching of the filaments to develop.  To illustrate the possible relation between $\gamma$ and the attainable fractal dimension $D$ of the resulting configuration, consider the following simple (non-fractal) special cases:  (1) spherical collapse to a point mass of dimension $D = 0$ is possible only if $\gamma < 4/3$; (2) cylindrical collapse to a line with a higher dimension $D = 1$ is possible only if $\gamma < 1$; and (3) collapse to a sheet with a still higher dimension $D = 2$ is possible only if $\gamma < 0$.  In these special cases, the maximum attainable dimension $D$ of the resulting configuration is related to $\gamma$ by $D = (4 - 3\gamma)/(2 - \gamma)$.  This relation can be derived in a more general way by noting that the Jeans length $\lambda_{\rm J}$ varies as $\rho^{(\gamma - 2)/2}$, while the Jeans mass $M_{\rm J}$ varies as $\rho^{(3\gamma - 4)/2}$; this implies that the number of Jeans-scale substructures that can form in a system of fixed total mass varies as $\rho^{(4 - 3\gamma)/2}$.  The number $N$ of bound substructures is then related to their size $L$ by $N \propto L^{-D}$, where $D = (4 - 3\gamma)/(2 - \gamma)$ and the exponent $D$ can be defined as the fractal dimension of the system.  If $\gamma \simeq 0.73$, as is expected for the low-density regime in star-forming clouds, the dimension predicted by this relation is $D \simeq 1.43$, which corresponds to a moderately complex fractal network.  Perhaps fortuitously, this dimension is similar to the fractal dimension $D \sim 1.4$ of the spatial distribution of newly formed stars that was found by Larson (1995) from an analysis of the clustering of young stars in Taurus.  Whatever may be the utility of such fractal descriptions, some collapse simulations do show the formation of at least moderately complex filamentary networks (e.g.\ Bonnell \& Bate 2002; Bate et al.\ 2003; Jappsen et al.\ 2005), so further study of the geometry of cloud collapse and fragmentation may be fruitful.

\section{DEPENDENCE ON OTHER PARAMETERS}  

\subsection{Metallicity}  

   If fragmentation and the IMF depend importantly on the thermal properties of star-forming clouds, they should depend on the abundances of the atoms, molecules, and dust grains that are responsible for the cooling, and hence on the overall metallicity.  Relatively few theoretical studies have considered the effect of metallicity on the thermal properties of collapsing clouds, but two that have done so are those of Low \& Lynden-Bell (1976) and Omukai (2000), with generally similar results.  Both studies considered the effect of a metallicity two orders of magnitude below solar, and both found, as might be expected, that this lower metallicity results in higher temperatures at all densities; the temperature is predicted to be higher by an order of magnitude or more at the lower densities, and by about 0.4 dex at the higher densities.  However, the temperature is also predicted to vary more strongly with density in the low-metallicity case, and there are now two temperature minima rather than one as in the solar-metallicity case.  These two minima, one associated with atomic cooling and the other with dust cooling, occur at lower and higher densities than the single minimum in the solar-metallicity case, and their associated Jeans masses are higher and lower, respectively.  Because there are now two temperature minima and two associated mass scales, it is not possible without further detailed collapse calculations to make simple predictions of the effect of metallicity on the overall mass scale of the stellar IMF.

   The lowest temperature is actually attained in the higher-density minimum that is associated with dust cooling, and as in the solar-metallicity case, it occurs when the gas becomes thermally coupled to the dust.  Because of the reduced dust abundance, thermal coupling requires a much higher density in the low-metallicity case, about $10^{-15}$\,g\,cm$^{-3}$.  The temperature at this point is predicted to be about 15 to 20~K, i.e.\ 3~to~4 times higher than the minimum temperature in the solar-metallicity case.  A similar result is obtained when the results of Larson (1973) are recalculated with a metallicity reduced by two orders of magnitude; this yields a minimum temperature of 20~K at a density of $3 \times 10^{-15}$\,g\,cm$^{-3}$.  Because both the temperature and the density at this new minimum point are considerably higher than in the solar-metallicity case, the associated Jeans mass is not greatly different, but it is somewhat reduced to about 0.05 to 0.1 solar masses.  If this mass scale is what primarily determines the peak mass of the IMF, the peak mass may actually {\it decrease} somewhat with decreasing metallicity, contrary to previous simple arguments that the mass scale should increase with decreasing metallicity (e.g.\ Larson 1998).  The possibility of forming very low-mass objects remains essentially the same as in the solar-metallicity case because the opacity-limit mass remains almost the same, i.e.\ about 0.01~M$_\odot$ or less, even with the reduced metallicity.  However, since there is now a secondary temperature minimum at a much lower density where the Jeans mass is several tens of solar masses, the net effect of a low metallicity on the IMF is not obvious, and it can only be determined by detailed numerical simulations like those of Jappsen et al.\ (2005).

   Low \& Lynden-Bell (1976) also considered the effect of having a reduced fraction of the heavy elements condensed into dust, and as a limiting case they considered the effect of having no dust at all, with a gas metallicity that is still two orders of magnitude below solar.  The primary temperature minimum associated with the thermal coupling of gas and dust then disappears and the temperature becomes much higher at these densities, leaving only the secondary minimum associated with atomic cooling, which occurs at a density of $10^{-19}$\,g\,cm$^{-3}$ where the temperature is predicted to be about 40~K.  The Jeans mass at this point is about 40~M$_\odot$, and the Jeans mass does not fall below 15~M$_\odot$ in the dust-free case until much higher densities and temperatures are reached where H$_2$ molecules control the thermal physics, as in the metal-free case (see below).  This suggests that the formation of low-mass stars is suppressed, or becomes very unlikely, in the absence of dust.  The dust plays a crucial role because only the thermal continuous emission from dust can provide enough cooling to reduce the Jeans mass below one solar mass at the intermediate densities where continuum optical depths are still low and fragmentation is still probable; line cooling by atoms and molecules is unimportant at these densities because the line optical depths become very large (Low \& Lynden-Bell 1976; Whitworth et al.\ 1998).  An important implication for cosmology may be that, whatever the overall abundance of heavy elements, the formation of low-mass stars does not become probable until significant amounts of dust have been created and dispersed.

   The completely metal-free case has been studied by many authors in order to understand the formation of the first stars in the universe (for a review see Bromm \& Larson 2004).  In this extreme case, the only available coolant is molecular hydrogen, and the predicted minimum temperature of about 200~K occurs at a density of the order of $10^{-20}$\,g\,cm$^{-3}$.  The Jeans mass at this point is of the order of $10^3$\,M$_\odot$, and the arguments given earlier suggest that this should lead to large typical masses for the first stars.  Although feedback effects are more important for massive stars than for low mass stars and may significantly reduce the efficiency of star formation, known feedback effects cannot prevent a metal-free accreting protostar from growing to a mass of at least 30~M$_\odot$ (Tan \& McKee 2004); a final mass of the order of 100~M$_\odot$ or more is possible, and a mass as large as 500~M$_\odot$ cannot be excluded (Bromm \& Loeb 2004).  The formation of any significant number of low-mass metal-free stars seems unlikely because very high compression would be required, for example in very thin dense filaments, but it cannot be completely excluded because the opacity-limit mass is still much lower than a solar mass even in this case (Omukai 2000; Omukai \& Yoshii 2003; Bromm \& Larson 2004).

   The effect of the first small amounts of heavy elements in providing enhanced cooling and thus possibly allowing the formation of the first significant number of low-mass stars has been studied by Bromm et al.\ (2001) and Bromm \& Loeb (2003), and they find that for the most important heavy-element coolant, carbon, a minimum abundance of 3.5 dex below solar is required in order for C$^+$ cooling to become more important than H$_2$ cooling.  Once C$^+$ cooling takes over from H$_2$ cooling, the gas rapidly cools to a temperature near the cosmic background temperature.  The resulting large drop in temperature, which may be by almost an order of magnitude, may then permit the formation of significant numbers of low-mass stars with a more normal IMF, provided that some dust has also been produced and dispersed, but detailed calculations are needed to explore this possibility further.

\subsection{Background radiation}  

   At any stage in the expansion of the universe, a minimum temperature is set by the cosmic background temperature $2.73(1 + z)$.  At present, this background temperature is below any predicted or observed cloud temperature, but not by a large factor; this means that the cosmic background radiation will become important for the thermal properties of star-forming clouds at not very large redshifts.  For example, at a redshift of $z = 5$ the cosmic background temperature is 16~K, high enough to make a considerable difference to the thermal properties and fragmentation of star-forming clouds as discussed in Sections 3 and~4.  If the cosmic background radiation sets a minimum cloud temperature of 16~K but the physics of star-forming clouds is otherwise unchanged, the transition from the initial cooling phase of collapse to the later isothermal phase occurs at a density of only $10^{-20}$\,g\,cm$^{-3}$, at which point the Jeans mass is about 20~M$_\odot$ and the Bonnor-Ebert mass is about 5~M$_\odot$.  These masses are an order of magnitude higher than those predicted in Section~4 even for a minimum temperature as high as 8~K, which is approximately the lowest temperature observed in present-day clouds.  According to the arguments given earlier, this should have an important effect on the stellar IMF, shifting the mass scale upward by at least an order of magnitude.  This upward shift in mass scale is even larger than the shift suggested to be possible by Larson (1998) on the basis of the simple assumption that pressures in present-day and high-redshift clouds are similar; this is because the transition from the cooling phase to the isothermal phase occurs at a lower pressure when the background temperature is higher.  An increase in the mass scale of the IMF at high redshifts can have many important consequences for galaxy evolution, as were discussed by Larson (1998).

   Another effect of the cosmic background radiation that could be important is a large increase in the opacity-limit mass, i.e.\ the minimum fragment mass set by the onset of high opacity at high densities.  Low \& Lynden-Bell (1976) noted that the opacity-limit mass increases strongly with the cosmic background temperature because of the strong increase of dust opacity with temperature; this minimum stellar mass increases to 0.07~M$_\odot$ at a redshift of 5 and to 1.2~M$_\odot$ at a redshift of 10.  Thus the formation of solar-mass stars may be completely suppressed at redshifts above 10.

   Another source of background radiation that can be important in star-forming clouds is the ambient far-infrared radiation from heated dust in regions of very active star formation.  Strong far-infrared emission is an observed property of such regions, and because optical depths are small at far-infrared wavelengths, this radiation can penetrate the dense cloud cores and heat the dust in them to temperatures significantly higher than would otherwise be predicted.  The effect of this ambient radiation on the thermal structure of dense cloud cores was calculated by Falgarone \& Puget (1985), and its possible implications for cloud fragmentation and the stellar IMF were discussed by Larson (1986), who speculated that the result might be a high-mass mode of star formation.  Falgarone \& Puget (1985) showed that the observed level of far-infrared radiation in star-forming giant molecular cloud complexes is sufficient to heat the dust throughout these complexes to a minimum temperature of 15 to 20~K.  As a result, the gas temperature is raised to at least 15~K at the high densities where the gas becomes thermally coupled to the dust, making the temperature at these densities a factor of 2 or 3 higher than would be predicted in the absence of dust heating (see Figure 1 of Larson 1986).  The gas temperature is predicted to have a dip at an intermediate density of $3 \times 10^{-20}$\,g\,cm$^{-3}$, where it reaches a minimum value of about 11~K.  At this point the Jeans mass is about 7~M$_\odot$ and the Bonnor-Ebert mass is about 1.5 M$_\odot$; these numbers are about 20 times larger than those predicted in Section~4 for a minimum temperature of 5~K at a density of $2 \times 10^{-18}$\,g\,cm$^{-3}$, and 3~times larger even than those predicted for a minimum temperature of 8~K reached at a density of $10^{-19}$\,g\,cm$^{-3}$.  This difference suggests that the far-infrared ambient radiation in regions of active star formation can have important effects on the thermal properties of star forming clouds and hence on the stellar IMF, possibly favoring a top-heavy IMF in such regions.  Such effects should be even larger in the more extreme circumstances existing in the vicinity of starbursts and AGNs.

   The above discussion assumes that the thermal properties of the gas at densities below that at which the gas becomes thermally coupled to the dust are not altered by the enhanced ambient radiation in star-forming regions, but this might not be the case if there is also enhanced gas heating by effects such as the photoelectric effect.  Falgarone \& Puget (1985) included enhanced photoelectric heating in their cloud models, but they found that its effect is limited by high optical depths at the relevant wavelengths, and their predicted temperature-density relation in the low-density regime is similar again to that found in the various studies mentioned in Section~4.  The temperature-density relation in the low-density regime may therefore not differ greatly in different regions, and the most important variable affecting the stellar IMF may just be the minimum dust temperature set by the background radiation field.  Larson (1985) noted that if only this minimum temperature varies, the predicted mass scale for fragmentation is quite sensitive to it, varying as $T_{\rm min}^{3.35}$.  This relation predicts, for example, that the mass scale for fragmentation increases by a factor of 10 for each factor of 2 increase in $T_{\rm min}$.  The numerical experiments of Jappsen et al.\ (2005) suggest a qualitatively similar but quantitatively somewhat weaker dependence of the typical fragment mass on minimum gas temperature, finding that the median fragment mass varies approximately as the 1.7 power of the minimum temperature or the -0.5 power of the associated density; this is still a very important effect, and it can have major consequences for the stellar IMF.

   Although the above discussion suggests that the ambient radiation in star-forming regions can be important for the mass scale of fragmentation, its net effect on the IMF of all the stars formed in a region may be reduced because the far-infrared radiation that heats the dust is produced only after a significant number of luminous stars have been formed, while it is possible that most of the low-mass stars have already formed by this time, especially if the low-mass stars tend to form before the more massive ones.  The effect of grain heating on the overall IMF may then be reduced, and the resulting IMF may not be very anomalous, or it may be anomalous only in local regions such as the cores of dense clusters.  The effect of heating by far-infrared radiation might be greater if a high level of ambient radiation were sustained for a long enough time for large numbers of stars of all masses to be formed; such a situation might occur, for example, in a starburst galactic nucleus where dense gas is retained and continues to form stars for a time much longer than the local dynamical time.

   Observationally, it remains unclear, even after extensive study, whether starburst regions have a systematically top-heavy IMF (Brandl \& Andersen 2005; Elmegreen 2005).  Most well-studied very luminous young clusters do not clearly show an anomalous IMF.  The strongest evidence for the existence of a top-heavy IMF may be provided by the Arches cluster, the best-studied luminous young cluster in the Galactic center region, a region of exceptionally intense massive star formation (Figer 2003, 2005; Stolte 2003; Stolte et al.\ 2002, 2005).  These authors find that the IMF of the Arches cluster is somewhat less steep than the Salpeter function at masses above 10~M$_\odot$, but a more important result may be the tentative finding of Stolte et al.\ (2005) that the IMF flattens and possibly begins to decline at masses below about 6~M$_\odot$.  This change in the slope of the IMF at 6~M$_\odot$, if confirmed by further studies over a larger area, would imply that the mass scale for star formation in this region is an order of magnitude higher than it is elsewhere in the Galaxy.  Other well-studied luminous young clusters that are outside the Galactic center region, such as NGC 3603, do not show any similar flattening of the IMF at intermediate masses (Stolte 2003).  The apparently anomalous IMF in the Galactic center region might result from the fact that this region contains the largest concentration of massive young stars and clusters in the Local Group (Figer 2003); the dust heating effect discussed above should therefore be much more important in this region than elsewhere.  The molecular clouds near the Galactic center are observed to be much warmer than molecular clouds elsewhere, typically about 70~K, and the dust temperature in this region is also quite high by the standards discussed above, about 40~K at the distance of the Arches cluster from the center (Morris \& Serabyn 1996).  Since both the gas temperature and the dust temperature are considerably increased by local heating effects, the situation is more complicated than the simple case discussed above where only the dust temperature is increased, but if the overall form of the temperature-density relation is similar to the form discussed earlier except for upward scaling (which will require more data to confirm), arguments like those given earlier suggest that a very top-heavy IMF will still result, with a mass scale that is higher than normal by at least an order of magnitude.  The IMF of the Arches cluster appears to show such an upward shift in mass scale by about an order of magnitude (Stolte et al.\ 2005), although as those authors note, further work is required before this result can be considered definitive.

   Even more extreme conditions may exist in starburst galactic nuclei, but no examples of such regions are presently accessible to detailed star count studies capable of yielding a direct determination of IMF.  Indirect evidence for a top-heavy IMF, or really a `bottom-light' IMF deficient in low-mass stars, has been found by Rieke et al.\ (1993) in the small starburst galaxy M82, and detailed studies of the `super star cluster' M82-F have also suggested such a bottom-light IMF in some clusters in M82 (Smith \& Gallagher 2001; McCrady, Graham \& Vacca 2005), although this evidence too is indirect and not conclusive.  If the inference of an IMF shifted to higher masses in M82 is indeed correct, and if starburst activity was more prevalent at earlier cosmological times, then the IMF may generally have been more top-heavy at earlier times, at least in the densest regions such as galactic nuclei.  Many important consequences would follow from an IMF that was systematically more top-heavy at earlier times, as were reviewed by Larson (1998).

\section{SUMMARY AND CONCLUSIONS}   

   Many observations have shown that the stellar IMF has a nearly universal standard form that is similar to the original Salpeter function at masses above a solar mass but flattens at lower masses and peaks (in logarithmic units) at a few tenths of a solar mass, declining again toward the lowest masses.  Similar and often indistinguishable results have been found in many diverse star-forming environments, and this suggests that the basic characteristics of the IMF, including its peak mass, are determined by relatively universal features of the physics of star formation.  Numerical simulations of cloud collapse and fragmentation show that the mass spectrum of the bound clumps that form is very sensitive to the assumed temperature-density relation, and this suggests that the IMF depends, at least for low-mass stars, on the detailed thermal physics of star-forming clouds.

   Observations and theory show that there are two important thermal regimes in star-forming clouds: a low-density regime controlled by atomic cooling in which the temperature decreases with increasing density, and a high density regime controlled by dust cooling in which the temperature increases slowly with increasing density.  Collapse simulations show that fragmentation is strongly favored in the low-density regime where the Jeans mass decreases strongly with increasing density, while much less fragmentation occurs in the high-density regime where the Jeans mass decreases much more slowly with increasing density.  This suggests that the Jeans mass at the transition point between these two regimes is a preferred mass scale for fragmentation, and that it may determine the peak mass of the IMF.  The Jeans mass at the predicted temperature minimum between the two regimes is about 0.3 solar masses, which is approximately the mass at which the IMF peaks.  This coincidence may be partly fortuitous because observations have not confirmed the very low predicted minimum temperature, but even if radiative heating effects raise the minimum temperature to 8~K, which is approximately the lowest value directly inferred from observations, the mass scale for fragmentation at the transition between the cooling phase of collapse and the later nearly isothermal phase is still of the order of one solar mass, and therefore it is still of relevance to the characteristic stellar mass.

   The characteristic stellar mass may thus be determined by fundamental physics, and more specifically by the conditions required for the thermal coupling of gas and dust.  The temperature when this occurs is predicted in most circumstances to be very low, less than 10~K, because the dust cools effectively by radiating in a continuum and not just in lines.  The density at which the thermal coupling occurs depends on the abundance of dust grains but is relatively high, of the order of $10^{-18}$\,g\,cm$^{-3}$ or more, because of the low number density of the grains.  The low temperature and high density at the point where thermal coupling occurs imply a small Jeans mass at this point, and this is what makes possible the formation of low-mass stars.  Numerical simulations support the hypothesis that this mass scale plays an important role in determining the peak mass of the stellar IMF.  Although other explanations of the peak mass have been proposed that are based on a characteristic turbulent or magnetic pressure, there is no apparent reason from fundamental physics for any preferred value for a turbulent or magnetic pressure; therefore an explanation based on the thermal physics may be the most promising way to account for the observed universality of the IMF.

   The strong sensitivity of the amount of fragmentation found in simulations to the temperature-density relation, and particularly to the difference between a decreasing and a slowly rising temperature, can be understood if a filamentary geometry plays an important role in fragmentation.  The isothermal case is a critical one for the collapse of a cylinder toward its axis: if the temperature decreases with increasing density, a cylinder can continue to collapse indefinitely and fragment into very small objects, while if the temperature begins to rise, this is no longer possible because the pressure then increases faster than gravity, slowing the collapse and causing the cylinder to break up into clumps.  Filamentary structure can result from dynamical effects such as shear, or from purely gravitational effects such as the amplification of departures from spherical symmetry during collapse.  The extent to which departures from spherical symmetry are amplified is limited by the rise of pressure, but if the temperature decreases sufficiently with increasing density, it may be possible not only for filaments to form but for fractal-like filamentary networks to develop.  Both simulations and observations often show the presence of filamentary networks in star-forming clouds.

   If the IMF of low-mass stars depends on the thermal physics of star-forming clouds, it may depend on metallicity, and it may also be influenced by ambient radiation fields such as the cosmic background radiation or the far-infrared radiation from heated dust in starburst regions.  The limited theoretical work that has been done on the effect of metallicity shows that the temperature-density relation becomes more complicated when the metallicity is reduced, and it is not clear without further detailed calculations whether this will lead to a systematic change in the IMF.  On the other hand, both the cosmic background radiation at high redshifts and the far-infrared radiation from heated dust in starburst regions will substantially raise the minimum temperature in star-forming clouds, and this significantly increases the Jeans mass at the transition point between the initial cooling phase of collapse and the later nearly isothermal phase.  As a result, there may be a significant upward shift in the mass scale of the IMF at early cosmological times or in starburst regions, and this would support the plausibility of a top-heavy IMF in these situations.  Further studies of the thermal properties of star-forming clouds in different circumstances and of their effects on the stellar IMF are needed to arrive at more definitive conclusions.

\label{lastpage}

\end{document}